# Surface-enhanced Gallium Arsenide photonic resonator with a quality factor of six million

Biswarup Guha,[1] Felix Marsault,[2] Fabian Cadiz,[3] Laurence Morgenroth,[4] Vladimir Ulin,[5] Vladimir Berkovitz,[5] Aristide Lemaître,[2] Carmen Gomez,[2] Alberto Amo,[2] Sylvain Combrié,[6] Bruno Gérard,[7] Giuseppe Leo,[1] and Ivan Favero[1]

[1]Matériaux et Phénomènes Quantiques, Université Paris Diderot, CNRS UMR 7162, Sorbonne Paris-Cité, 10 rue Alice Domon et Léonie Duquet, 75013 Paris, France
[2]Laboratoire de Photonique et de Nanostructures, CNRS UPR 20, Route de Nozay, 91460 Marcoussis, France
[3]Laboratoire de Physique de la Matière Condensée, Ecole Polytechnique, CNRS UMR 7643, Route de Saclay, 91128 Palaiseau, France
[4]Institut d'Electronique, de Microélectronique et de Nanotechnologie, UMR CNRS 8520, Avenue Poincaré, 59652 Villeneuve d'Ascq
[5] A. F. Ioffe Physico-Technical Institute, 194021 Saint Petersburg, Russia
[6]Thales Research and Technology, 1 avenue Augustin Fresnel, 91767 Palaiseau, France
[7]III-V Lab, 1 Avenue Augustin Fresnel, 91767 Palaiseau, France

**Gallium Arsenide and related compound semiconductors lie at the heart of optoelectronics and integrated laser technologies. Shaped at the micro and nano-scale, they allow strong interaction with quantum dots and quantum wells, and promise to result in stunning devices. However gallium arsenide optical structures presently exhibit lower performances than their silicon-based counterparts, notably in nanophotonics where the surface plays a chief role. Here we report on advanced surface control of miniature gallium arsenide optical resonators, using two distinct techniques that produce permanent results. One leads to extend the lifetime of free-carriers and enhance luminescence, while the other strongly reduces surface absorption originating from mid-gap states and enables ultra-low optical dissipation devices. With such surface control, the quality factor of wavelength-sized optical disk resonators is observed to rise up to six million at telecom wavelength, greatly surpassing previous realizations and opening new prospects for Gallium Arsenide nanophotonics.**

The ability to strongly confine light is crucial in today's photonic sciences, with applications that span from quantum investigation of light-matter interactions to the advancement of new optical sources, sensors and detectors. In dielectrics, just like in intrinsic semiconductors, light cannot be steadily localized below half the wavelength in the material [1]. At telecom wavelength, this leads to ultimate optical mode volumes ranging from a few micron cubes down to a fraction of a micron cube, reached for the most refractive semiconductors. This strong spatial confinement has its temporal counterpart, quantified by the optical quality factor Q. Over the last decade, Qs in the mid-$10^6$ and up to nine million have been progressively obtained with miniature silicon cavities [2,3]. Even though Gallium Arsenide material (GaAs) is just as refractive as silicon, the best equivalent photonic structures fabricated out of GaAs have so far saturated an order of magnitude below, with Qs in the mid-$10^5$ [4-9], and very little progress lately. Equaling the performances of silicon nanophotonics with GaAs would open many prospects, considering the material direct bandgap, its strong quadratic non-linearity, and its natural compliance with optically active elements like quantum dots and quantum wells. In quantum dot cavity-QED, the coherence of manipulated states and the cadence of single photons emission would for example increase. Nonlinearities at the single photon level [10,11], enabling generation of non-classical states of light, would become accessible in GaAs cavities, using the mere optical non-linearity of the material [12-15], or tailored interactions with a mechanical element [16-18]. Extending the carrier lifetime in GaAs etched structures would allow the fabrication of submicron polariton resonators in the strong coupling regime [19]. The performances of miniature GaAs-based lasers [20-24] would also improve, leading to reduced power consumption and parasitic heating, and better spectral purity. In a recent work, we investigated in great detail the case of high-Q (mid-$10^5$) GaAs whispering gallery resonators, and showed that surface absorption was a major source of optical loss in these devices [25]. Here we demonstrate advanced surface control, and break open the lock associated to such surface dissipation. With two distinct techniques, wet nitridation and atomic layer deposition (ALD), both producing permanent results, we slow-down the relaxation dynamics of free-carriers and mitigate the surface absorption currently pinning the performances of devices. As a result, the luminescence of miniature GaAs photonic resonators increases and their optical Q rises up to six million at telecom wavelength, equaling the state-of-the-art of silicon photonics.

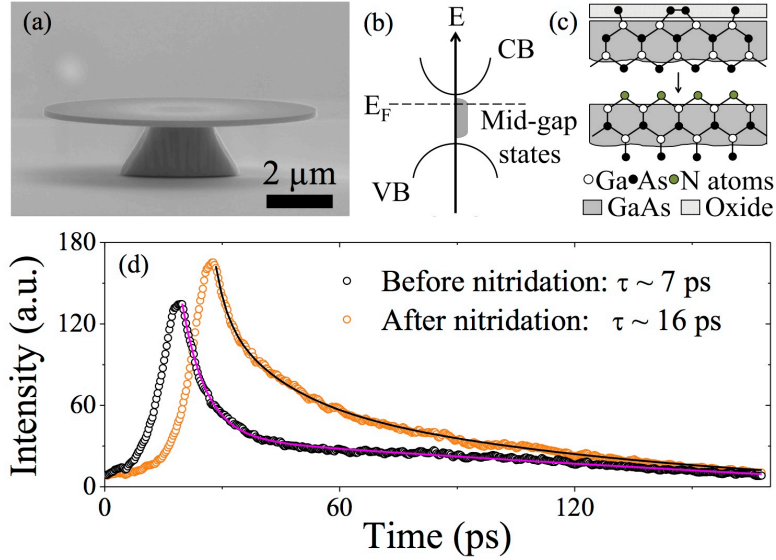

Fig. 1. Wet nitridation of GaAs disk resonators. (a) Side-view electron micrograph of a GaAs disk sitting atop a central AlGaAs pedestal. (b) Representation of the intra-gap density of states associated to the resonator's surface, with arbitrary pinning of the Fermi level $E_F$. (c) Schematics of the nitridation process, where an arsenic-rich surface oxide is replaced by a stable layer of nitrogen atoms. (d) Time-resolved photoluminescence of a single GaAs disk before and after nitridation.

Our disk resonators are isolated from the substrate through an Aluminum Gallium Arsenide (AlGaAs) pedestal, as visible in Fig. 1(a). They are fabricated out of a GaAs (200 nm)/$Al_{0.8}Ga_{0.2}As$ (1800 nm)/GaAs epitaxial wafer, by electron beam lithography, non-selective wet etch and final selective under-etch in hydrofluoric acid [7,25]. Two epitaxial wafers (A and B) will be employed in this work, with the same nominal structure. With a typical diameter of a few microns, the fabricated disks support whispering gallery modes (WGMs) in the telecom range, with theoretical Q superior to $10^8$, determined by bending losses. In practice, our best resonators have to date reached $Q=5.10^5$, and we established that the reconstruction layer at the resonator's surface was playing a central part in this discrepancy [25]. Indeed the electronic states associated to this layer can reside in the gap of the material, and the surface chemical nature also pins the Fermi level. The situation is represented in Fig. 1(b), showing that the resonator with its reconstructed surface can absorb photons having energy below the material bandgap. By controlling the structural and chemical nature of the surface, one can tune the density of mid-gap states and the Fermi level, producing variable effects on optical absorption, but on other properties as well, like the carrier-dynamics and luminescence.

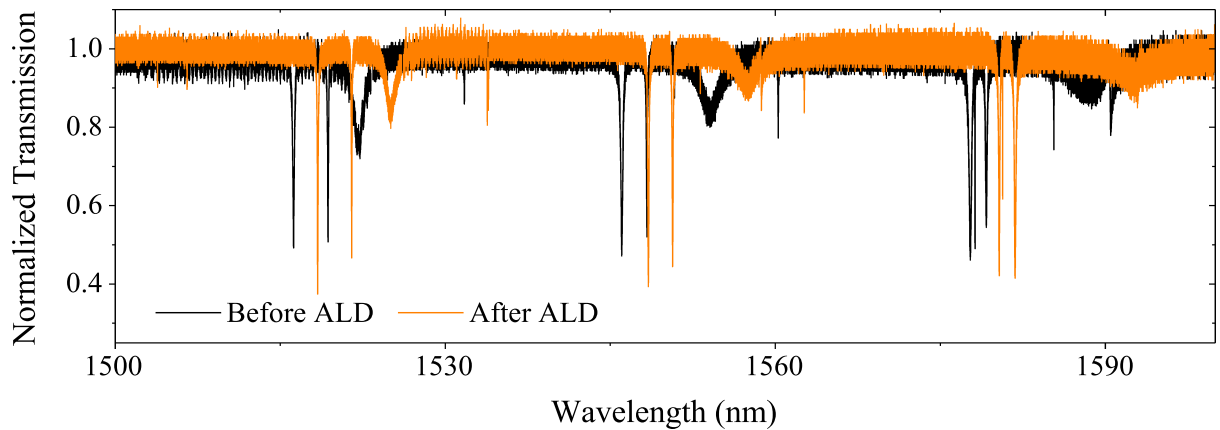

Fig. 2. Effect of ALD surface treatment on the optical spectrum of a GaAs disk resonator. After ALD, WGM resonances are red-shifted and narrowed, and new thin resonances appear with low contrast.

We first investigate a process of wet nitridation using a hydrazine ($H_2S_4$) solution, which leads to the formation of a stable GaN layer on the surface (Fig. 1(c)). Surface nitridation has been shown to enhance the photoluminescence of a piece of bulk GaAs under continuous-wave pumping [26], but it is generally accompanied by a detrimental surface etching. In order to circumvent this problem and preserve the morphology of our miniature resonators, we intentionally lower the PH of the employed solution and bring the parasitic etching to a negligible amount (see supplementary material). We observe a marked effect of the resulting surface treatment in room-temperature time-resolved photoluminescence experiments performed on individual disks. Such an experiment is reported in Fig. 1(d) and reveals a rich dynamics, with a decay fitted by the sum of a rapid, an intermediate and a slow exponential function. Here the intermediate decay corresponds to the relaxation of the population of carriers that give rise to the luminescence [27,28]. This decay is systematically slowed-down by the surface nitridation, with a decay time $\tau$ increasing from 7±3 ps to 16±4 ps after nitridation for the investigated disks. Under continuous pumping, the luminescence of a single disk experiences a 20-fold enhancement, with no visible degradation over a month. However, these pronounced effects on the carrier-dynamics do not show clear counterparts in terms of optical absorption. Our measurement of GaAs disks WGMS by fiber-taper optical spectroscopy [29] showed neither improvement of optical Qs nor reduction of absorption effects after nitridation of the surface. As we will see below, this situation changes drastically with other passivation methods.

We now move to a second surface control technique, consisting in Atomic Layer Deposition (ALD) of alumina ($Al_2O_3$), which produces extremely thin, uniform and conformal barrier layers on the resonator's surface (see supplementary material). Fig. 2 shows the optical spectrum of a single GaAs disk resonator before and after such ALD treatment. The spectra are obtained by fiber-taper optical spectroscopy, and reveal a series of fine dips corresponding to WGM resonances of the resonator [29,7], spreading over a 100nm wavelength span. These resonances spectrally bunch in distinct groups, separated by the free spectral range between WGMs of adjacent azimuthal numbers [30]. After ALD, the overall spectral structure of the spectrum is preserved, while a global red shift is observed, consistent with the refractive response of the deposited alumina layer. The optical resonances are narrowed, while their contrast gets trendily increased, pointing towards a reduction of intrinsic optical losses of the resonators. A finer understanding of this effect is obtained by performing similar experiments at higher optical power, as we show hereafter.

Fig. 3 shows selected WGM resonances of a GaAs disk resonator, measured before and after ALD, at low and high optical power. The spectral identification of a given WGM before and after treatment is made possible by the analysis of a full spectrum of the type shown in Fig. 2. At low power, the WGM resonance in Fig.3 (a) and (b) adopts a Lorentzian shape, and gets narrowed by a factor 2.3 after ALD treatment, with a constant contrast of 40%. In the usual coupled-modes theory (CMT) of fiber-taper coupling experiments [7,29], this implies a 2.3-fold reduction of the intrinsic loss rate of the WGM $\kappa_{in}$, meaning reduction of the optical absorption rate $\kappa_{abs}$ by at least the same factor. At high power, the Lorentzian resonance is distorted by thermo-optical effects, with a wavelength drag that corresponds to the elevation of the disk temperature produced by absorption [25]. In Fig.3 (c) and (d), the WGM resonance experiences an important reduction of this drag after ALD treatment, by a factor of about 7. Since at constant contrast the absorbed optical power scales like $\kappa_{abs}/\kappa_{in}$ [7], the optical absorption is reduced by at least a factor 7 to produce such effect. These two examples prove that an important mitigation of absorption can be obtained by ALD. The intensity of the effect shall of course depend on the overlap of the considered WGM with surface absorption sites. In our most performing GaAs disks, with ultra-smooth surfaces, the optical Q of the best WGMs was limited by surface absorption [25], hence ALD passivation should lead stunning results in these cases.

Fig. 4 reports more systematic fiber-taper optical measurements performed at low power on an ALD-passivated GaAs disk of radius 5 micron. The upper panel of Fig. 4 is a WGM spectrum obtained in the under-coupled regime, and reveals a fine-structure doublet typical of high-Q situation [7,29]. The loaded Q for each component of the doublet is $2.8 \times 10^6$ and $3 \times 10^6$ respectively, already an order of magnitude above results obtained on disks without surface treatment [29]. In order to disentangle properly the dissipative effects associated to the presence of the fiber-taper, we vary the gap distance between the fiber taper and resonator. The results are shown in the lower panel of Fig. 4. At large gap-distance, the line width of the WGM resonance converges to an intrinsic line width of 0.26 ±0.03 pm, corresponding to an intrinsic Q of 5.9 (±0.6) $\times 10^6$. We observe similar values on several WGMs and resonators having experienced same surface treatment. These results represent a 10-fold improvement with respect to prior state-of-the-art in GaAs photonic cavities, and clearly underline the potential of proper surface control. Interestingly, these ultra-high-Q WGMs coupled to a fiber taper do not behave as expected from the standard CMT, in contrast to smaller-Q cases [29]. At 500 nm of gap-distance, Fig. 4 shows that the WGM resonance line width is increased by a factor two by fiber taper loading, while the contrast hardly reaches 20% instead of the 100% expected from regular CMT in this case.

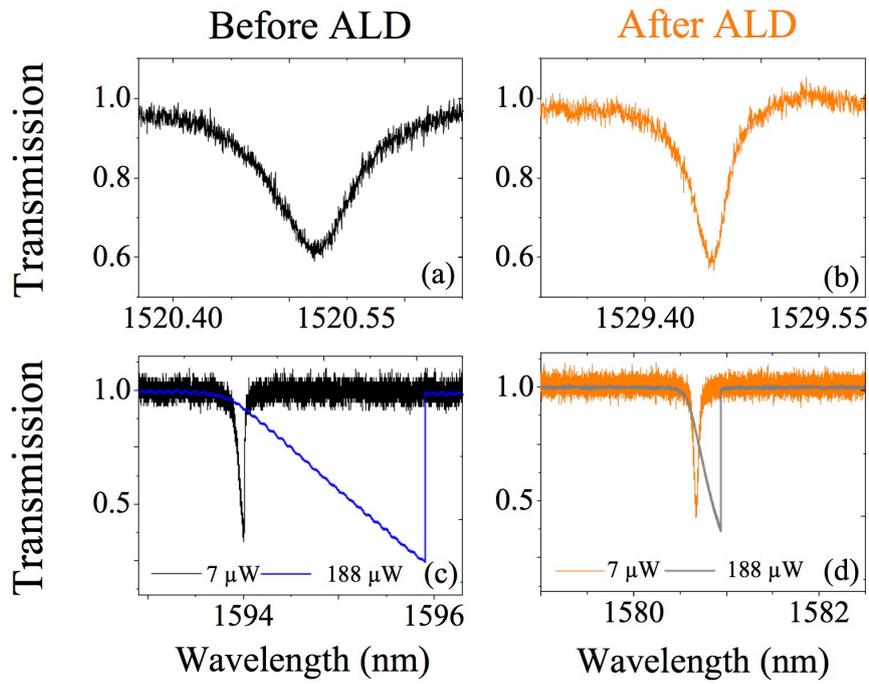

Fig. 3. Reduction of optical absorption by ALD passivation. (a) A selected WGM resonance of a 5 μm radius GaAs disk, measured at low optical power before ALD, with FWHM of 84 pm. (b) Same resonance after ALD, with FWHM of 36 pm. The resonance is red-shifted by the deposition of 20 nm of alumina. (c and d) Another WGM resonance, recorded on a larger wavelength span, showing thermo-optical response before (c) and after (d) ALD with 30 nm of alumina. The blue shift is due to chemical pre-treatment with ammonia prior to ALD [25].

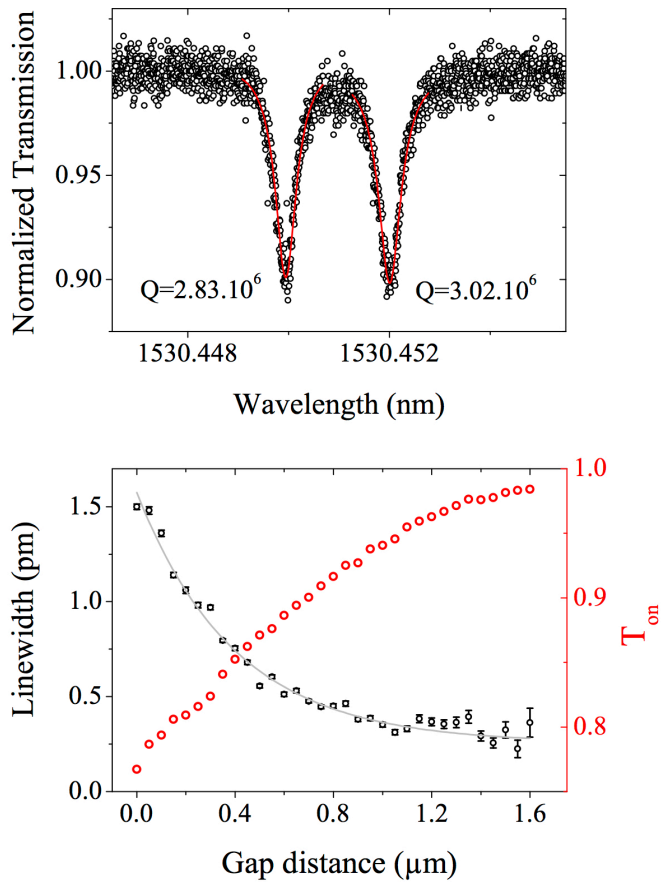

Fig. 4. GaAs disk with a Q of six million. (a) Fiber-taper optical spectrum of an ultra-high Q WGM measured on a 5 μm radius GaAs disk passivated by ALD (under-coupled regime). (b) Line width and resonant normalized transmission of the WGM resonance as a function of the fiber to resonator gap-distance, showing intrinsic Q of six million.

Finally, we investigate how the ALD passivation method and its beneficial effect on Q are impacted by the WGM mode profile, the thickness of the ALD-deposited alumina, and the intrinsic quality of the employed GaAs material. In Fig. 5 we show the measured Q enhancement obtained by ALD for WGMs of different radial order p [30], for different deposited thicknesses, and for the two epitaxial wafers A and B employed in this study. Qs are analyzed and compared before and after ALD, in a loaded configuration producing constant optical resonance contrast of about 50%. In such configuration, the measured enhancement factor varies between 1 and 5. The first trend is that whatever the deposited thickness and the employed substrate, the Q enhancement improves as p decreases, corresponding to a localization of electromagnetic energy closer to the surface. This further supports that the main source of optical dissipation in high-Q GaAs WGMs originates from surfaces [25]. The second conclusion reached from Fig. 5 is the absence of marked improvement when the deposited alumina thickness increases from 5 to 10, 20 and 30 nm. This indicates that a layer of 5 nm suffices to obtain beneficial features. The last observation is the qualitatively different behavior of wafers A and wafer B, which correspond to two distinct hetero-epitaxial growth of the same nominal structure. Both wafers are expected to posses a residual p-doping in the $10^{14}/cm^3$ range, however for the same WGM and same ALD thickness, the Q enhancement is more pronounced for wafer B, pointing towards less residual bulk absorption. This illustrates that with Qs above the million, usually insignificant variations of the crystal may now matter.

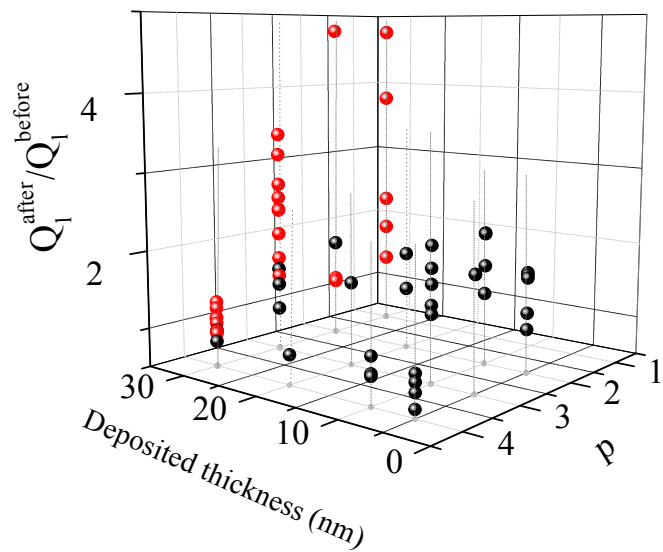

*Fig. 5. Variations of the Q-enhancement by ALD. The enhancement of loaded quality factor $Q_l$ is measured at constant contrast of the optical resonance as a function of the WGM radial number p, of the deposited alumina thickness, and for two distinct wafers A (black) and B (red).*

In conclusion, novel surface control techniques enable bringing GaAs photonic cavities to a new stage. Even though we not did reach yet the theoretical limit set by bending losses, the presented WGM resonators already attain optical Q in the million range and up to six million. The results reported here show no degradation upon time, with a stability observed over several months under ambient laboratory conditions. These advances should be beneficial to multiple fields of research utilizing GaAs photonics, like solid-state cavity QED, nonlinear optics, optoelectronics, and optomechanics [6,7]. Interestingly, the two surface treatments employed here provide contradictory trends: one (nitridation) slows down carrier relaxation but does not impact surface absorption at 1.55μm of wavelength, while the other (alumina ALD) strongly reduces surface absorption but produces variable effects on carrier relaxation, with basically no effect in the GaAs disks studied here, and appreciable effects in photonic crystals [31]. These observations call for a better understanding of the exact relation between microscopic aspects like spectral density of surface states, surface pinning of the Fermi level, or surface chemical state, and macroscopic effective properties like the optical and transport responses of the resonator at various frequencies. In the context of optomechanics, for metrological and quantum applications, the impact of these microscopic aspects on mechanical dissipation [32,33] will also be central. From these different examples, it becomes clear that nanoscale resonators will require advanced surface physics and engineering to reveal their potential in full.

*We acknowledge the European Research Council (ERC) (Ganoms StG project, 20111012, GA306664) and the ANR-DGA (Octopuss project). We thank Daniel Paget and David Parrain for first discussions and attempts towards surface control of disks.*

1. J. B. Khurgin, "How to deal with the loss in plasmonics and metamaterials", Nature Nanotechnology **10**, 2–6 (2015).
2. Y. Taguchi, Y. Takahashi, Y. Sato, T. Asano and S. Noda, "Statistical studies of photonic heterostructure nanocavities with an average Q factor of three million," Optics Express **19**(12), 11916–11921 (2011).
3. H. Sekoguchi, Y.Takahashi, T. Asano, and S. Noda, "Photonic crystal nanocavity with a *Q*-factor of ~9 million," Optics Express **22**(1), 916-924 (2014).
4. C. P. Michael, K. Srinivasan, T. J. Johnson, O. Painter, K. H. Lee, K. Hennessy, H. Kim, and E. Hu, "Wavelength- and material-dependent absorption in gaas and algaas microcavities," Appl. Phys. Lett. **90**(5), 051108 (2007).
5. S. Combrié, A. De Rossi, Q. V. Tran, and H. Benisty, "GaAs photonic crystal cavity with ultrahigh Q: microwatt nonlinearity at 1.55 μm," Opt. Lett. **33**(16), 1908–1910 (2008).
6. L. Ding, C. Baker, P. Senellart, A. Lemaître, S. Ducci, G. Leo, and I. Favero,"High frequency GaAs nano-optomechanical disk resonator," Phys. Rev. Lett. **105**(26), 263903 (2010).
7. J. C. L. Ding, C. Baker, A. Andronico, D. Parrain, P. Senellart, A. Lemaître, S. Ducci, G. Leo, and I. Favero, "Gallium arsenide disk optomechanical resonators," in Handbook of Optical Microcavities (PanStanford, 2014).
8. C. Arnold, V. Loo, A. Lemaître, I. Sagnes, O. Krebs, P. Voisin, P. Senellart, L. Lanco, "Optical bistability in a quantum dots/micropillar device with a quality factor exceeding 200 000,"Appl. Phys. Lett. **100**, 111111 (2012).
9. S. Reitzenstein, C. Hofmann, A. Gorbunov, M. Strauss, S.H. Kwon, C. Schneider, A. Löffler, S. Höfling, M. Kamp and A. Forchel, "AlAs∕GaAs micropillar cavities with quality factors exceeding 150.000", App. Phys. Lett. **90**, 251109 (2007).
10. A. Faraon, I. Fushman, D. Englund, N. Stoltz, P. Petroff and J. Vuckovic,"Coherent generation of nonclassical light on a chip via photon-induced tunneling and blockade", Nature Physics **4**, 859-863 (2008).
11. V. Loo, C. Arnold, O. Gazzano, A. Lemaître, I. Sagnes, O. Krebs, P. Voisin, P. Senellart, and L. Lanco, "Optical Nonlinearity for Few-Photon Pulses on a Quantum Dot-Pillar Cavity Device", Phys. Rev. Lett. **109**, 166806 (2012).
12. M. Bamba, A. Imamoğlu, I. Carusotto, and C. Ciuti, "Origin of strong photon antibunching in weakly nonlinear photonic molecules", Phys. Rev. A **83**, 021802(R) (2011).
13. A. Andronico, I. Favero, and G. Leo, "Difference frequency generation in GaAs microdisks", Optics Letters **33**(18), 2026-2028 (2008).
14. P. S. Kuo, J. Bravo-Abad and G. S. Solomon," Second-harmonic generation using quasi-phasematching in a GaAs whispering-gallery-mode microcavity", Nature Comm **5**, 3109 (2013).
15. S. Mariani, A. Andronico, A. Lemaître, I. Favero, S. Ducci, and G. Leo, "Second-harmonic generation in AlGaAs microdisks in the telecom range", Opt. Lett. **39**, 3062 (2014).
16. A. Nunnenkamp, K. Børkje, and S. M. Girvin," Single-Photon Optomechanics", Phys. Rev. Lett. **107**, 063602 (2011).
17. P. Rabl, "Photon Blockade Effect in Optomechanical Systems", Phys. Rev. Lett. **107**, 063601 (2011).
18. J. Restrepo, C. Ciuti, and I. Favero, "Single-Polariton Optomechanics", Phys. Rev. Lett. **112**, 013601 (2014).
19. D. Bajoni, P. Senellart, E. Wertz, I. Sagnes, A. Miard, A. Lemaitre and J. Bloch. "Polariton Laser Using Single Micropillar GaAs-GaAlAs Semiconductor Cavities", Phys. Rev. Lett. **100**, 47401 (2008).
20. F. Albert, T. Braun, T. Heidel, C. Schneider, S. Reitzenstein, S. Höfling, L. Worschech, and A. Forchel, "Whispering gallery mode lasing in electrically driven quantum dot micropillars", Appl. Phys. Lett **97**, 101108, (2010).
21. A. Tandaechanurat, S. Ishida, D. Guimard, M. Nomura, S. Iwamoto and Y. Arakawa, "Lasing oscillation in a three-dimensional photonic crystal nanocavity with a complete bandgap", Nature Photonics **5**, 91–94 (2011).
22. B. Ellis, M. A. Mayer, G. Shambat, T. Sarmiento, J. Harris, E. E. Haller and J. Vučković, "Ultralow-threshold electrically pumped quantum-dot photonic-crystal nanocavity laser", Nature Photonics **5**, 297–300 (2011).
23. D. Saxena, S. Mokkapati, P. Parkinson, N. Jiang, Q. Gao, H. Hoe Tan and C . Jagadish, "Optically pumped room-temperature GaAs nanowire lasers", Nature Photonics **7**, 963–968 (2013).
24. J. Tatebayashi, S. Kako, J. Ho, Y. Ota, S. Iwamoto and Y. Arakawa, "Room-temperature lasing in a single nanowire with quantum dots", Nature Photonics **9**, 501–505 (2015).
25. D. Parrain, C. Baker, G. Wang, B. Guha, E. Gil-Santos, A. Lemaître, P. Senellart, G. Leo, S. Ducci, and I. Favero," Origin of optical losses in gallium arsenide disk whispering gallery resonators", Optics Express **23**(15), 19656-19672 (2015).
26. V. L. Berkovits, D. Paget, A. N. Karpenko, V.P. Ulin, O.E. Tereshchenko, Appl. Phys. Lett. **90**, 022104 (2007).
27. R. Eccleston, R. Strobel, W. W. Rühle, J. Kuhl, B. F. Feuerbacher, and K. Ploog, "Exciton dynamics in a GaAs quantum well", Phys. Rev. B 44, 1395 (1999).
28. A. Amo, M. D. Martín, L. Viña, A. I. Toropov, and K. S. Zhuravlev, "Interplay of exciton and electron-hole plasma recombination on the photoluminescence dynamics in bulk GaAs", Phys. Rev. B **73**, 035205 (2006).
29. L. Ding, P. Senellart, A. Lemaitre, S. Ducci, G. Leo, and I. Favero. "GaAs micro-nanodisks probed by a looped fiber taper for optomechanics applications, Proc. SPIE 7712, 771211 (2010).
30. A. Andronico, X. Caillet, I. Favero, S. Ducci, V. Berger, G. Leo. "Semiconductor microcavities for enhanced nonlinear optics interactions", J. Europ. Opt. Soc. Rap. Public. **3**, 08030 (2008).
31. G. Moille, S. Combrié, L. Morgenroth, G. Lehoucq, F. Neuilly, B. Hu, D. Decoster and A. de Rossi, "Integrated all-optical switch with 10 ps time resolution enabled by ALD", Laser Photonics Rev **1**-11 (2016).
32. D. T. Nguyen, C. Baker, W. Hease, S. Sejil, P. Senellart, A. Lemaître, S. Ducci, G. Leo et I. Favero. "Ultrahigh Q-Frequency product for optomechanical disk resonators with a mechanical shield", Appl. Phys. Lett. **103**, 241112 (2013).

33. D. T. Nguyen, W. Hease, C. Baker, E. Gil-Santos, P. Senellart, A. Lemaître, S. Ducci, G. Leo, and I. Favero. "Improved optomechanical disk resonator sitting on a pedestal mechanical shield", New Journal of Physics **17**, 023016 (2015).

**Supplements :**

**Wet nitridation of GaAs disk optical resonators.** Thin layers of GaN are promising candidates for GaAs surface passivation thanks to the high stability of GaN bonds. In addition, because of their high electronegativity, nitrogen atoms bonded onto surface gallium atoms should not lead to electronic states within the gap of GaAs. To form a thin nitride film, a wet chemical nitridation procedure has been developed [34,35]. It consists in treating the GaAs crystal in a highly alkaline (pH=12) hydrazine ($N_2H_4$) solution with a small amount (0.01M) of sodium sulfide ($Na_2S$) added. The latter removes surface arsenic atoms under the form of soluble thioarsenic acid $H_3AsS_3$. The surface nitridation proceeds through dissociative adsorption of hydrazine molecules on the opened gallium atoms. The process stops after the formation of a continuous monolayer of GaN [35]. The formed nitride monolayer protects the crystal surface against oxidation. Unfortunately, the process also comes with surface microetching, which can affect the morphology of miniature structures [35,36]. This parasitic etching originates from the interaction of hydroxyl anions $OH^-$ with surface gallium atoms. To circumvent this problem, here we intentionally decreased the pH of the hydrazine–sulfide solution down to a value of 8.5. The preparation of the solution consists in two steps: firstly, we obtain a hydrazine buffer solution by adding anhydrous hydrazine-dihydrochloride ($N_2H_4\times2HCl$) to hydrazine-hydrate, until reaching a target pH value. Secondly, $Na_2S$ is introduced into the solution up to a concentration of 0.01 M, leading to a concentration of $OH^-$ anions lowered by 4 orders of magnitude and a negligible parasitic etching [37].

Before nitridation, the samples supporting GaAs disk resonators were rinsed in methylene chloride ($CH_2Cl_2$) and acetone, and immersed in concentrated $NH_4OH$ solution for a few minutes before rinsing with water. After drying they were immersed in the low alkaline hydrazine –sulfide solution for 10 minutes at 80°C. The samples were subsequently rinsed in deionized water.

**Time-resolved photoluminescence of GaAs disks.** Photoluminescence experiments are performed using a pulsed Ti:sapphire laser delivering 3 ps pulses at a repetition rate of 82MHz. A microscope objective 50x (NA=0.65) is used both to focus the laser on a 2 µm spot and to collect the emission, which is focused on a monochromator slit with a lens of 300 mm focal length. The emitted signal is dispersed in the monocromator and time resolved using a streak camera, with a time resolution of 8 ps. A broadband longpass filter with a cutoff wavelength of 780 nm prevents the excitation laser at 775 nm from reaching the detector. A schematics of these experiments is provided below.

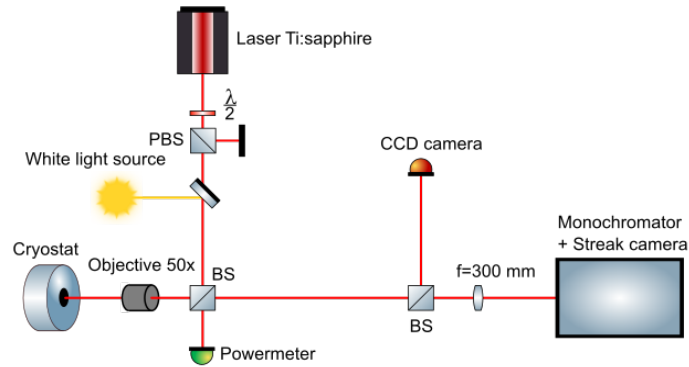

Fig. S1. Time-resolved photoluminescence experiments. PBS: Polarization Beam-Splitter. BS- Beam-splitter. The cryostat temperature is kept at room temperature in this work.

**Atomic Layer Deposition on GaAs disk resonators.** In this work, the ALD coating was deposited using a TFS 200 reactor (Beneq Oy). The samples received hydrogen plasma pre-treatment in order to improve the interfacial quality, before using a plasma-enhanced atomic layer deposition (PE-ALD), in order to deposit a layer of alumina. ALD itself consists in a series of ALD cycles, which are made up of a sequence of four pulses in rapid succession: alternating pulses of precursor, here trimethyl-aluminium $Al(CH_3)_3$, and reactant, here $O_2$ plasma. Between precursor and reactant pulses the chamber is purged by a 300sccm flow of argon gas. Depositions were performed at a temperature of 300 °C. The number of precursor cycles for each deposition was calculated using a growth rate per cycle of ~ 0.1 nm/cycle for alumina.

34. V. L. Berkovits, V. P. Ulin, M. Losurdo, P. Capezzuto, G. Bruno, G. Perna, V. Capozzi, Appl. Phys. Lett. 80, 3739 (2002).
35. V. L. Berkovits, V. P. Ulin, O. E. Tereshchenko, D. Paget, A.C.H. Rowe, P. Chiaradia, B.P. Doyle, S. Nannarone, J. Electrochem. Soc. 158, D127 (2011).
36. V. L. Berkovits, L. Masson, I. V. Makarenko, V.P Ulin. Appl. Surf. Sci. 254, 8023 (2008).
37. P. A. Alekseev, M. S. Dunaevskiy, V. P. Ulin, T. V. Lvova, D. O. Filatov, A.V. Nezhdanov, A. I. Mashin, V. L. Berkovits, Nano Lett. 15, 63 (2015).